\documentclass[11pt]{article}

\usepackage[text={17.1cm,24.6cm},centering]{geometry} 
\usepackage[utf8]{inputenc}
\usepackage{authblk}
\usepackage{blkarray}
\usepackage{cite}

\usepackage{amsfonts,amsmath,amssymb,amsthm}
\usepackage{latexsym,mathrsfs,mathtools,bm}
\usepackage{graphicx,subcaption,epsfig,caption,float,xcolor}
\usepackage{enumitem}
\usepackage{chngcntr}

\usepackage[normalem]{ulem}
\usepackage{cancel}

\usepackage{xcolor}

\usepackage{tikz}
\usetikzlibrary{calc}

\usepackage{bookmark}

\newtheorem{theo}{Theorem}[section]

\newtheorem*{theo*}{Theorem}

\newtheorem{remax}[theo]{Remark}
\newenvironment{rema}
  {\pushQED{\qed}\remax}
  {\popQED\endremax}
  
\theoremstyle{definition}

\newtheorem*{term*}{Notation/Terminology}

\usepackage{geometry}
\numberwithin{equation}{section}

\usepackage{ulem}
\usepackage{xcolor}

\newcommand\redsout{\bgroup\markoverwith{\textcolor{red}{\rule[0.5ex]{2pt}{0.4pt}}}\ULon}

\title{\bf Perfect state transfer in inhomogeneous XX model\\ of $q$-Racah type}
\renewcommand*{\Affilfont}{\normalsize\small}

\author[1,a]{Nicolas Cramp\'e\,}
\author[2,b]{Simon Lafrance\,}
\author[2,c]{Charles Robillard\,}
\author[2,3,d]{Luc Vinet\,\vspace{.5em}}
\affil[1]{\textit{CNRS -- Universit\'e de Montr\'eal CRM - CNRS, Montr\'eal (Qu\'ebec), H3C 3J7, Canada}\vspace{.9em}}
\affil[2]{\textit{Centre de Recherches Math\'ematiques, Universit\'e de Montr\'eal, P.O. Box 6128,
\newline\vspace{.9em}
 Centre-ville Station, Montr\'eal (Qu\'ebec), H3C 3J7, Canada}}
\affil[3]{\textit{IVADO, Montr\'eal (Qu\'ebec), H2S 3H1, Canada \newline\vspace{.9em}}}

{
 \makeatletter
 \renewcommand\AB@affilsepx{: \protect\Affilfont}
 \makeatother
 \makeatletter
 \renewcommand\AB@affilsepx{, \protect\Affilfont}
 \makeatother
 \affil[a]{crampe1977@gmail.com}
 \affil[b]{simon.lafrance.3@umontreal.ca}
  \affil[c]{charles.robillard.1@umontreal.ca}
  \affil[d]{luc.vinet@umontreal.ca}
}
\date{\today}

\begin{document}

\maketitle

\bigskip\bigskip 

\begin{center}
\begin{minipage}{12cm}
\begin{center}
{\bf Abstract} 
\\
\end{center}
New exactly solvable one-dimensional XX spin chain models that exhibit perfect state transfer are defined. These models have inhomogeneous couplings and magnetic fields determined from the three-term recurrence relations satisfied by the $q$-Racah and para $q$-Racah polynomials. Due to this connection with orthogonal polynomials, the one-excitation sector can be solved analytically. This allows us to provide explicit sets of conditions on the polynomial parameters that guarantee the occurrence of perfect state transfer across these spin chains.
    \end{minipage}
\end{center}

\medskip

\begin{center}
\begin{minipage}{8cm}
\end{minipage}
\end{center}

\clearpage
\newpage

\section{Introduction}

The interplay between exactly solvable physical models and the mathematical framework of orthogonal polynomials and special functions is a cornerstone of mathematical physics. This paper illustrates this relationship by investigating non-uniform spin chains featuring Perfect State Transfer (PST) that are designed using orthogonal polynomials from the $q$-Askey scheme \cite{KoekoekLeskyetal2010}. 

The ability to transfer a quantum state from one location to another is fundamental to quantum information and its processing, as it underpins communication between quantum processors and the development of quantum algorithms. One-dimensional spin chains offer interesting means for building the quantum wires necessary for this transport. These models are attractive because the transport is mediated autonomously by the dynamics of the chain, avoiding the need for external control. The ultimate goal is perfect state transfer where the input state will be found at the output location with probability one at a specified time.

It has been shown that a PST can be achieved by carefully engineering the couplings between spins of the XX spin chains with nearest-neighbor inhomogeneous couplings \cite{albanese2004mirror}, \cite{christandl2005perfect}, \cite{Kay10}, \cite{VZ12}. In this case, the one-excitation dynamics is governed by a Jacobi matrix (a tridiagonal Hermitian matrix) and the conditions for PST boil down to persymmetry (or mirror symmetry) of the Jacobi matrix and to the requirement that consecutive eigenvalues differ by odd integers (up to a scale factor). Thus, determining an XX spin chain with PST reduces to an inverse spectral problem: constructing a Jacobi matrix that possesses a specific set of eigenvalues. The profound connection between Jacobi matrices and the three-term recurrence relations defining orthogonal polynomials provides the crucial tool for solving this problem. When the underlying orthogonal polynomials enjoy a full characterization, an analytic chain is obtained. A handful of these instances have been found and a review of the models of that type known to harbour PST (as well as Fractional Revival) has been written \cite{bosse2017coherent}. The references to many of the original articles can be found in this survey. It might be worth stressing that the actual design of such devices can be helped from the knowledge of the specifications through explicit formulas, moreover the non-uniformity of the couplings and magnetic fields does not seem to pose overwhelming engineering challenges \cite{perez2013coherent}, \cite{chapman2016experimental}. Finding exactly solvable models proceeds from a long tradition in theoretical physics and expanding the set of such models has always proven worthwhile; this report goes into this direction.

In this paper, we study spin chains associated with the $q$-Racah polynomials and determine the conditions for  these chains to possess PST. The two cases that emerge when the persymmetry condition is imposed are studied in detail. In the second case, attention needs to be paid  to the fact that a divergence arises in the coupling constants but a careful scaling of the parameters allows us to overcome this problem. This leads to chains associated with the para $q$-Racah polynomials defined in \cite{Lemay18}. The conditions on the eigenvalues are then solved for these two cases, and it is thus found that numerous models with PST arise, associated with integers satisfying some inequalities. Let us mention that the analysis of the $q$-Racah cases recovers one special case that had been given in \cite{VZ12}. Furthermore, it should be stressed that the identification of these new families of spin chains with PST involve picking as the case may be, specific values of $q$. This should be put in contrast to the work carried out in \cite{chakrabarti2010quantum} and \cite{jafarov2010quantum}. In the spirit of the present study, these authors undertook to determine spin 
chains with PST that correspond to the orthogonal polynomials of the Askey 
scheme. Jafarov and Van der Jeugt consider the families of the $q$-Askey tableau 
and concluded that PST was possible only for one type of $q$-Krawtchouk and not 
for chains associated with other $q$-orthogonal polynomials. While this would 
seem to contradict the results presented here, this is not so because the authors 
of \cite{jafarov2010quantum} kept $q$ a generic rational number and did not consider the possibility of fixing this parameter to special values that are moreover irrational in the cases that we identify.

The plan of this paper is as follows. Section \ref{sec:PST} contains the definition of the XX model and recalls the  necessary and sufficient conditions for PST to occur, namely the persymmetry condition and the constraints on the differences between two consecutive eigenvalues.  Section \ref{sec:qR} introduces the 
$q$-Racah polynomials and identifies two parameter regimes for which the 
persymmetry property is satisfied. In the first regime, we prove that the model 
exhibits PST provided that the remaining free parameters are expressed in terms 
of three integers satisfying certain inequalities. Section \ref{sec:para} 
describes the second regime and its connection with the para $q$-Racah 
polynomials. Section \ref{sec:outlook} concludes the paper with remarks and open 
problems.

\section{Perfect state transfer and orthogonal polynomials \label{sec:PST}}

Consider the open quantum XX spin chain with $N+1$ sites labeled from  $0$ to $N$ and described by the following Hamiltonian
\begin{align}\label{eq:H}
    H= \frac{1}{2}\sum_{\ell=0}^{N-1} J_{\ell+1} \left(\sigma^x_\ell \sigma^x_{\ell+1}
   + \sigma^y_\ell \sigma^y_{\ell+1}\right) +\frac12\sum_{\ell=0}^{N} b_\ell(\sigma^z_\ell+1)\,,
\end{align}
where $J_\ell$, real parameters, are the coupling constants between the sites 
$\ell-1$ and $\ell$, $b_\ell$ are the strengths of the magnetic field at the sites $\ell$, and the symbols $\sigma^x$, $\sigma^y$, $\sigma^z$ stand for the
Pauli matrices.

Let us denote by $|n\rangle$ ($n=0,1,\dots, N$) the vectors with one excitation, \textit{i.e.}, with only one spin down:
\begin{align}
  |n\rangle= \underbrace{\begin{pmatrix}
      1\\0
  \end{pmatrix} \otimes\cdots\otimes \begin{pmatrix}
      1\\0
  \end{pmatrix}  }_{n}\otimes \begin{pmatrix}
      0\\1
  \end{pmatrix}\otimes \underbrace{\begin{pmatrix}
      1\\0
  \end{pmatrix}  \otimes\cdots\otimes \begin{pmatrix}
      1\\0
  \end{pmatrix}  }_{N-n}.
\end{align}
A perfect state transfer (PST) between the sites $0$ and $N$ corresponds to the fact that the state $|0\rangle$ is transferred to the state $|N\rangle$ after a certain time $T$ with respect to the dynamics given by $H$.
The goal is to find parameters $J_\ell$ and $b_\ell$ such that this phenomena can occur.

The subspace spanned by the vectors $|n\rangle$ ($n=0,1,\dots,N$) is isomorphic to a vector space $V$ of dimension $N+1$. In the canonical basis $\{e_n\ | \ n=0,1,\dots,N\}$ of $V$, the Hamiltonian $H$ reduces to a $(N+1)\times(N+1)$ tridiagonal matrix
\begin{align}
    \mathcal{H}=\begin{pmatrix}
        b_0 & J_1 &  \\
        J_1 & b_1 & J_2\\
        &\ddots& \ddots&\ddots\\
        && J_{N-1} & b_{N-1} & J_{N} \\
    &&& J_{N} & b_{N} 
    \end{pmatrix}\,.
\end{align}
In this setting,  PST occurs if there exists a time $T$ such that 
\begin{align}
    e^{iT\mathcal{H}}e_0 =e^{i\phi}e_N,
\end{align}
where $\phi$ is a real parameter.

We denote by $\epsilon_x$ ($x=0,1,\dots,N$) the set of ordered eigenvalues of $\mathcal{H}$: $\epsilon_0< \epsilon_1< \cdots <  \epsilon_N$. 
It is known that the PST property is equivalent to the two conditions \cite{Kay10,VZ12}:
\begin{itemize}
    \item the eigenvalues satisfy, for $x=0,1,\dots N-1$,
    \begin{align}\label{eq:cond1}
        \epsilon_{x+1} - \epsilon_x = \frac{\pi}{T} M_x\,,
    \end{align}
where $M_x$ are positive odd integers;
\item the Hamiltonian $\mathcal{H}$ is persymmetric, \textit{i.e.}, $R\mathcal{H}R = \mathcal{H}$ where $R$ is the reflection matrix
\begin{align}
    R=\left(\begin{matrix}
        &&&1\\
        &&1\\
        &\scalebox{-1}[1]{$\ddots$}\\
        1
    \end{matrix}\right)\,.
\end{align}
\end{itemize}
The eigenvectors associated to the eigenvalue $\epsilon_x$ of $\mathcal{H}$ are $
\psi_x=\big(p_0(x),p_1(x),\cdots,p_N(x)\big)^t\,,
$
such that the eigenvalue problem $\mathcal{H}\psi_x=\epsilon_x \psi_x$ is equivalent to, for $n=0,1,\dots,N$,
\begin{align}
   J_{n} p_{n-1}(x)+ b_n p_n(x)+  J_{n+1} p_{n+1}(x)  =\epsilon_x p_n(x)\,,
\end{align}
where, by convention $J_{0}=J_{N+1}=0$. The previous relation defines, by recurrence, polynomials $p_n(x)$ which form an orthogonal family by Favard's theorem \cite{Chiara}. Therefore,  the study of the orthogonal polynomials is the natural setting in which to obtain perfect state transfer in XX models. This approach has been used with success for different polynomials: the symmetric Krawtchouk polynomials and a special case of the $q$-Racah polynomials \cite{VZ12}, the para-Krawtchouk polynomials \cite{VZ12c} and the dual $-1$ Hahn polynomials \cite{VZ12b}, see \cite{bosse2017coherent} a compendium.

In this paper, we propose to use this approach starting from the $q$-Racah polynomials or a $q$-generalization of the para Racah polynomials introduced in \cite{Lemay18} in order to obtain new models with perfect state transfer. Let us mention \cite{Scherer21,Scherer21b} where specializations of these polynomials have been used to construct classical models with perfect state transfer.

\section{$q$-Racah polynomials and the first model \label{sec:qR}}

Let us recall that the $q$-Racah polynomials $R_n$ satisfy the following  recurrence relation (see \textit{e.g.} \cite{KoekoekLeskyetal2010}):
\begin{align}
  \lambda_x R_n(x) =A_n R_{n+1}(x) +B_n R_n(x) +C_n R_{n-1}(x) \,,
\end{align}
where $
    \lambda_x=q^{-x}+\gamma\delta q^{x+1}\,,
$
and
\begin{align}
&A_n=\frac{(1-\alpha q^{n+1})(1-\alpha\beta q^{n+1})(1-\beta\delta q^{n+1})(1-\gamma q^{n+1})}{(1-\alpha\beta q^{2n+1})(1-\alpha\beta q^{2n+2})}\,,\\
     &C_n=\frac{q(1-q^{n})(1-\beta q^{n})(\gamma-\alpha\beta q^{n})(\delta-\alpha q^{n})}{(1-\alpha\beta q^{2n})(1-\alpha\beta q^{2n+1})}\,,\qquad B_n=-A_n-C_n+1+\gamma\delta q\,.
\end{align}
We choose $0<q<1$ and $\gamma=q^{-N-1}$.
These polynomials, after suitable normalizations, allow us to diagonalize $\mathcal{H}$ with
\begin{align}
    J_n=\mu\sqrt{A_{n-1}C_{n}} \,,\qquad b_n=\mu B_n\,,
\end{align}
and $\mu$ a free parameter.
For these choices, the eigenvalues of $\mathcal{H}$ become
\begin{align}\label{eq:eigen}
    \epsilon_x=\mu( q^{-x} + \delta q^{x-N})\,.
\end{align}

We want to find the constraints on the parameters $q$, $\alpha$, $\beta$, $\delta$ and $\mu$ such that the Hamiltonian possesses the PST property. For the persymmetry conditions \textit{i.e.},
\begin{align}\label{eq:cons1}
    b_n=b_{N-n}\,,\qquad J_n=J_{N-n+1}\,,
\end{align}
    there are two following  solutions:
    \begin{itemize}
        \item[(i)] $\alpha\beta q^{N+1}=-1$\,,\quad $\delta=\alpha^2 q^{N+1}$\,; 
        \item[(ii)] $\alpha\beta q^{N+1}=1$\,.
    \end{itemize}
If the conditions in the item (i) of the previous list are satisfied, we can use $\beta=-q^{-N-1}/\alpha$ and $\delta=\alpha^2 q^{N+1}$ to simplify $b_n$ and $J_n$:
\begin{align}
 &   b_n=\mu \frac{(1 + q)( 1+q^{N+1})(1+\alpha^2 q^{N+1} )}{(q^{N+1-n} + q^{n})(q^{n+1}+ q^{N-n})}\,,\label{eq:bn1}\\
  &  J_n^2=\mu^2 \frac{(1-\alpha^2 q^{2n})(1-\alpha^2 q^{2N+2-2n} )(1-q^{2n})(1-q^{2N+2-2n} )}{( q^{n-1}+q^{N+1-n} )( q^{n}+q^{N+1-n})^2(q^{n} + q^{N-n})}\,.\label{eq:Jn1}
\end{align}
We can verify directly that these parameters satisfy the persymmetry conditions \ref{eq:cons1}.
It remains to prove that the eigenvalues
$
    \epsilon_x=\mu (q^{-x}+\alpha^2 q^{x+1})\,,
$
satisfy condition \eqref{eq:cond1}.
These constraints can be solved for $x=0$ and $x=1$ in terms of $\mu$ and $\alpha^2$:
\begin{align}\label{eq:amu}
    \alpha^2 = \frac{M_0-M_1q }{q^3(M_0q - M_1)}\,,\quad \mu =\frac{\pi}{T} \frac{q^2(M_1-M_0 q)}{(1+q )(1-q)^2}.
\end{align}
Injecting these results in condition \eqref{eq:cond1}, one obtains, for $x=0,1,\dots N-1$,
\begin{align}\label{eq:M0M1Mx}
   \frac{q^{x}-q^{-x}}{q-q^{-1}}M_1- \frac{q^{x-1}-q^{-x+1}}{q-q^{-1}}M_0=M_x\,.
\end{align}
Evidently, this last relation is satisfied for $x=0,1$ and it can be simplified using the Chebyshev polynomials of the second kind $U_n$ defined recursively by:
\begin{align}\label{eq:recu}
    U_{-1}(x)=0\,,\quad U_0(x)=1\,,\quad U_{n+1}(x)=2x U_n(x)-U_{n-1}(x)\quad (n=0,1,\dots)\,,
\end{align}
and satisfying, for $x=0,1,\dots$, 
\begin{align}
    U_{x}\left(\frac{q+q^{-1}}{2}\right)=\frac{q^{x+1}-q^{-x-1}}{q-q^{-1}}\,.
\end{align}
Using these results, the constraints \eqref{eq:M0M1Mx} can be written as follows, for $x=2,3,\dots N-1$, 
\begin{align}\label{eq:MM}
   U_{x-1}\left(\frac{q+q^{-1}}{2}\right) M_1-  U_{x-2}\left(\frac{q+q^{-1}}{2}\right)M_0=M_x\,.
\end{align}
If we choose $q+q^{-1}$ to be an even integer 
\begin{align}\label{eq:vq}
    q+q^{-1}=2m\,,
\end{align}
then the l.h.s. of \eqref{eq:MM} becomes an integer (the Chebyshev polynomials are polynomials with integer coefficients). 
Since $q$ is chosen such that $0<q<1$, we have $m>1$ and we must choose the following solution for $q$:
\begin{align}\label{eq:vqm}
    q=m-\sqrt{m^2-1}\,,
\end{align}
between both possible solutions of \eqref{eq:vq}.
It remains to prove that the left-hand side of \eqref{eq:MM} is odd and positive.
Using the recurrence relations \eqref{eq:recu} satisfied by $U_n$, one can show that $U_n(m)$ is even when $n$ is odd and $U_n(m)$ is odd when $n$ is even. Therefore, for any  integer $m$ and any odd integers $M_0$ and $M_1$, we can prove that
\begin{align}
    U_{x-1}(m)M_1-U_{x-2}(m)M_0\,,
\end{align}
is an odd integer for $x=2,3,\dots, N-1$.

The integers $M_x$ for $x=2,3,\dots N-1$ are positive if and only if
\begin{align}
   \frac{1-q^{2x}}{1-q^{2x-2}}M_1>q M_0 \,.
\end{align}
We have used the expression \eqref{eq:M0M1Mx} of $M_x$. Knowing that $0<q<1$, these inequalities are satisfied for all $x$ if it satisfied for $x=N-1$. 

Let us summarize the result. 
First choose an integer $m>1$ which gives $q$ from \eqref{eq:vqm} and, for any choice of positive odd integers $M_0$ and $M_1$ satisfying
\begin{align}\label{eq:cond}
   \frac{1-q^{2N-2}}{1-q^{2N-4}}M_1>qM_0\,,
\end{align}
let compute $\alpha^2$ and $\mu$ from \eqref{eq:amu}. Therefore, the model described by the Hamiltonian \eqref{eq:H}, with the couplings given by \eqref{eq:bn1},\eqref{eq:Jn1}, exhibits PST.

Let us remark that the coefficients $J_n^2$ are positive if $\alpha^2q^{2n}<1$, for $1\leq n\leq N$, which provide additional constraints given by $M_0(1-q^2)/(M_0q-M_1)<q$. If we demand that the constraints \eqref{eq:cond} be valid for any $N$, this leads to $M_1>qM_0$ and, in this case, the coefficients $J_n^2$ are automatically positive.

\begin{rema}
  The example treated in \cite{VZ12} corresponds to a particular choice of $M_0$ and $M_1$. For example, for $N$ even, it is obtained for
  \begin{align}
      M_0=U_{N/2-1}\left(\frac{q+q^{-1}}{2}\right)-U_{N/2-2}\left(\frac{q+q^{-1}}{2}\right)\,,\\
      M_1=U_{N/2-2}\left(\frac{q+q^{-1}}{2}\right)-U_{N/2-3}\left(\frac{q+q^{-1}}{2}\right)\,.
  \end{align}
  Injecting these values in \eqref{eq:amu}, one obtains $
      \alpha^2=-q^{-N-1}$ and $\mu=-\frac{\pi}{T}\frac{q^{N/2}}{q-q^{-1}}\,,
$  and for the energies \begin{align}
      \epsilon_x=\frac{\pi}{T}\frac{q^{-x+N/2}-q^{x-N/2}}{q-q^{-1}}\,.
  \end{align}
  Let us emphasize that the parameters $b_n$ of the Hamiltonian vanish for this particular choice.
\end{rema}

\section{Second model and para $q$-Racah polynomials \label{sec:para}}

Let us now focus on the second case when $\alpha\beta q^{N+1}=1$.
Under these constraints, a pole appears in the coupling constant $J_n$. Therefore, at first glance, it seems impossible to use these models. However, this problem has been overcome in \cite{Lemay18} where a suitable definition of the parameters followed by a limit allows one to deal with this pole. This approach leads to the definition of the para $q$-Racah polynomials. We refer to \cite{Lemay18} for the details of this construction. We provide here only the recurrence coefficients of these polynomials needed to construct the models.

\subsection{Recurrence relations of the para $q$-Racah polynomials}

The recurrence relations of the para $q$-Racah polynomials
depend on the parity of $N$. 
For the odd case $N=2j+1$,  
the coefficients appearing in this relation are\cite{Lemay18}:
\begin{align}
  &  A_n=       \frac{(1 - \alpha\beta q^n)(\beta - \alpha q^{n-j} )(1 - q^{n-2j-1})}{
\alpha\beta(1 - q^{2n-2j-1})(1 + q^{n-j} ) }\,,
\quad C_n=\frac{(1-q^n)(\alpha-\beta q^{n-j-1})(\alpha\beta-q^{n-2j-1})}{\alpha\beta(1-q^{2n-2j-1})(1+q^{n-j-1})}\,.
\end{align}
The Hamiltonian
$\mathcal{H}$ defined with these coefficients ($\mu$ a free parameter)
\begin{align}
    J_n=\mu\sqrt{A_{n-1}C_{n}} \,,\qquad b_n=\mu (\alpha+\alpha^{-1}-A_n-C_n)\,,
\end{align}
is already persymmetric and has the following eigenvalues:
\begin{align}
  &  \epsilon_{2x}=\mu(\alpha q^x +\alpha^{-1} q^{-x})\,,\qquad  \epsilon_{2x+1}=\mu(\beta q^x +\beta^{-1} q^{-x}) \qquad (x=0,1,\dots j)   \,.
\end{align}
Let us emphasize that the results in \cite{Lemay18} are a bit more general but we restrict ourselves to the persymmetric case.

For the even case $N=2j$, the coefficients appearing in the recurrence relation are\cite{Lemay18}:
\begin{align}
  &  A_n=       \frac{(1 - \alpha\beta q^n)(\beta - \alpha q^{n-j+1} )(1 - q^{n-2j})}{
\alpha\beta(1 - q^{2n-2j+1})(1 + q^{n-j} ) }\,,\quad C_n=\frac{(1-q^n)(\alpha-\beta q^{n-j-1})(\alpha\beta-q^{n-2j})}{\alpha\beta(1-q^{2n-2j-1})(1+q^{n-j})}\,.
\end{align}
In this case, the Hamiltonian is also persymmetric
and has the following eigenvalues:
\begin{align}
  &  \epsilon_{2x}=\mu(\alpha q^x +\alpha^{-1} q^{-x}), \text{ for }0\leq x\leq j\,,\quad \epsilon_{2x+1}=\mu(\beta q^x +\beta^{-1} q^{-x}), \text{ for }0\leq x\leq  j-1  \,.
\end{align}

\subsection{Constraints on the eigenvalues}

The Hamiltonian is persymmetric so we only need to prove that the eigenvalues satisfy the constraints  \eqref{eq:cond1} in order to have PST.

The differences of two successive eigenvalues for these models (independently of the parity of $N$) are given by 
    \begin{align}\label{eq:diff}
&    \epsilon_{2x+1}-\epsilon_{2x}=\mu(\alpha-\beta)\frac{1-\alpha\beta q^{2x}}{\alpha\beta q^x}\,,\qquad   \epsilon_{2x+2}-\epsilon_{2x+1}=\mu(\alpha q-\beta)\frac{\alpha\beta q^{2x+1}-1}{\alpha\beta q^{x+1}}\,.
\end{align}
Demanding that these differences satisfied the constraints \eqref{eq:cond1} for $x=0,1,2$ fixes $\alpha$, $\beta$ and $\mu$ in terms of the odd positive integers $M_0$, $M_1$ and $M_2$. The explicit expressions are not provided because their form is involved and they are not essential for the following arguments.

Using these relations, one can show that the differences \eqref{eq:diff} can be written in terms of Chebyshev polynomials:
\begin{align}
  &    \epsilon_{2x+1}-\epsilon_{2x}=\frac{\pi}{T}\left(U_{x-1}\left(\frac{q+q^{-1}}{2}\right)M_2-U_{x-2}\left(\frac{q+q^{-1}}{2}\right)M_0\right)\,,\\
 &    \epsilon_{2x+2}-\epsilon_{2x+1}=\frac{\pi}{T}\frac{M_1}{M_0+M_2}\left(U_{2x}\left(\frac{q^{1/2}+q^{-1/2}}{2}\right)M_2-U_{2x-2}\left(\frac{q^{1/2}+q^{-1/2}}{2}\right)M_0\right)\,.  
\end{align}
Let the ratio $M_1/(M_0+M_2)$ be an odd integer and $q$ be given by
\begin{align}
    q^{1/2}=m-\sqrt{m^2-1}\,\qquad m=2,3,\dots\,.
\end{align}
In this case, one gets $(q^{1/2}+q^{-1/2})/2=m$ and $(q+q^{-1})/2=2m^2-1$ which proves that the constraints \eqref{eq:cond1} are satisfied with $M_x$ odd integers. The proof follows the same structure as the previous argument. These integers are positive if the following constraint is satisfied:
\begin{align}\label{eq:cond2}
   q^2\frac{1-q^{N-1}}{1-q^{N-3}}M_2>M_0\,.
\end{align}
The preceding results show that a model exhibiting PST can be defined for any choice of integer $m>1$, and for any positive odd integers $M_0$, $M_1$, and $M_2$, provided that condition \eqref{eq:cond2} is satisfied and the ratio $M_1/(M_0+M_2)$ is an odd integer.

\section{Outlook \label{sec:outlook}}

We have introduced analytically solvable XX spin chain models that exhibit perfect state transfer, achieved through inhomogeneous nearest-neighbor couplings. The models were constructed using the recurrence relations of the $q$-Racah polynomials and the para $q$-Racah polynomials. We anticipate that these models will provide novel examples demonstrating phenomena such as almost perfect state transfer \cite{VZ12d} or fractional revival \cite{GVZ16}. Additionally, the well-studied multivariate generalizations of the $q$-Racah polynomials \cite{Tratnik,GR,CFGR} are expected to be useful for studying PST in multi-dimensional systems, extending the work done with multivariate dual Hahn polynomials in \cite{Miki}.\\

\noindent
\textbf{Acknowledgments: } N.~Cramp\'e is partially supported by the international research project AAPT of the CNRS. L.~Vinet is funded in part by a Discovery Grant from the Natural Sciences and Engineering Research Council (NSERC) of Canada. Simon Lafrance and Charles Robillard have received support from this grant and S.L. gratefully acknowledges receiving a CRM-ISM Undergraduate Research Scholarship.

\bibliographystyle{utphys}
\bibliography{PST}

\end{document}